\documentstyle[onecolumn,rotating,psfig]{mnv2}
\newif\ifAMStwofonts
\AMStwofontstrue
\def\kms{\rm{km s^{-1}}}
\def\mpc{h^{-1}{\rm Mpc}}
\def\Mpc{\rm Mpc}
\def\cA{${\cal A}$}
\def\cB{${\cal B}$}
\def\cC{${\cal C}$}

\def\calH{{$\cal H$}\,}
\def\cetal{Songaila \& Cowie}
\def\retal{Rauch et. al.}
\def\gsim{~\rlap{$>$}{\lower 1.0ex\hbox{$\sim$}}}
\def\dhi{n_{_{\rm HI}}}
\def\dhhi{\hat n_{_{\rm HI}}}

\def\nhi{\dhi}

\def\vp{v_{\rm p}}

\def\ltsim{\lower.5ex\hbox{$\; \buildrel < \over \sim \;$}}
\def\gtsim{\lower.5ex\hbox{$\; \buildrel > \over \sim \;$}}
\def\ltsim{\lower.5ex\hbox{$\; \buildrel < \over \sim \;$}}
\def\gtsim{\lower.5ex\hbox{$\; \buildrel > \over \sim \;$}}

\def\vx{{\bf x}}

\def\mgh#1{{}}
\def\adi#1{{}}

\def\s{\,{\rm s}}
\def\K{\,{\rm K}}
\def\kms{\mbox{km\,s$^{-1}$}}

\def\dd{\,{\rm d}}

\newcommand{\op}{Ly$\alpha$\ }
\newcommand{\hi}{\mbox{H{\scriptsize I}}}

\newcommand{\num}{\nu_{\rm c}}
\newcommand{\ttau}{\tilde \tau}
\newcommand{\ttaum}{\ttau_{\rm c}}
\newcommand{\ttaut}{\ttau_{\rm t}}

\begin{document}
\title[]{The amplitude of mass density fluctuations at $z\approx 3.25$ from the 
\op forest of Q1422+231}

\author[Nusser \& Haehnelt] { Adi Nusser$^1$ and Martin Haehnelt$^2$ \\
$^1$ Physics Department, The Technion-Israel Institute of Technology\\ 
$^2$ Max-Planck-Institut f\"ur Astrophysik, Karl-Schwarzschild-Str. 1, 
85740 Garching}
\maketitle

\begin{abstract}
The real-space optical depth distribution along the line of sight to
the QSO Q1422+231 is recovered from two HIRES spectra using a modified
version of the inversion method proposed by Nusser \& Haehnelt (1999).
The first two moments of the truncated optical depth distribution are
used to constrain the density fluctuation amplitude of the
intergalactic medium (IGM) assuming that the IGM is photoionized by a
metagalactic UV background and obeys a temperature-density relation.
The fluctuation amplitude and the power law index $ \alpha$ of the
relation between gas and neutral hydrogen (\hi ) density are
degenerate. The $rms$ of the IGM density at $z\approx 3.25$ estimated
from the first spectrum is $\sigma = \sqrt{\exp{[(1.8 \pm
0.27)^2/\alpha^2]}-1}$, with $1.56 <\alpha <2$ for plausible
reionization histories.  This corresponds to $0.9 \la \sigma \la 2.1$
with $\sigma(\alpha =1.7)= 1.44\pm 0.3 $. The values obtained from
the second spectrum are higher by $\approx 20 \%$.  If the IGM density
traces the dark matter (DM) as suggested by numerical simulations we
have measured the fluctuation amplitude of the DM density at an
effective Jeans scale of about a hundred to two hundred (comoving)
kpc. For CDM-like power spectra the amplitude of dark matter
fluctuations on these small scales depends on the cosmological density
parameter $\Omega$. For power spectra normalized to reproduce the
space density of present-day clusters and with a slope parameter of
$\Gamma=0.21$ consistent with the observed galaxy power spectrum, the
inferred $\Omega$ can be expressed as: $\Omega= 0.61
\left(\alpha/1.7\right)^{1.3}\left(x_{_{\rm J}} /0.62\right)^{-0.6}$
for a flat universe, and $\Omega= 0.91\left( \alpha/1.7\right)^{1.3}
\left(x_{_{\rm J}}/0.62\right)^{-0.7}$ for a $\lambda=0$ universe.
$x_{_{\rm J}}$ is the effective Jeans scale in (comoving)
$\mpc$. Based on a suit of detailed mock spectra the 1-$\sigma$ error
is $\approx 25 \%$. The estimates increase with increasing
$\Gamma$. For the second spectrum we obtain $15\%$ lower values.
\end{abstract}

\section{Introduction} 
The Lyman forest in QSO absorption spectra is now generally believed
to be caused by absorption by large-scale  \hi\
density fluctuations of moderate amplitude in a warm photo-ionized
intergalactic medium (IGM). This is substantially different from the
old cloud picture for the \op forest (see Rauch 1998 for a review). It
is mainly sustained by cosmological hydrodynamical simulations which
successfully reproduce the absorption features in QSO spectra (Cen et
al.~1994; Petitjean, M\"ucket \& Kates 1995; Zhang, Anninos \& Norman
1995; Hernquist et al.~1996; Miralda-Escud\'e et al.~1996, Theuns et
al. 1998). The picture had, however, been suggested before these
numerical simulations became available (Bond, Szalay \& Silk 1988; Bi,
B\"orner \& Chu 1992). Important results of the simulations are a
tight correlation between the \hi\ and the dark matter distribution
(on scales larger then the Jeans length of the IGM) and a simple gas
temperature-density relation which depends only on the reionization
history of the Universe (Hui \& Gnedin 1997, Haehnelt \& Steinmetz
1998). As a consequence, the \op forest can be an important probe for
the distribution of dark matter and its evolution over a wide redshift
range (c.f. Weinberg 1999 for a review) on scales smaller than those
accessible by the galaxy distribution or the lensing effects of
foreground large scale structures on background galaxies
(c.f. Bartelmann \& Schneider 1999).

For example, Gnedin \& Hui (1998) showed that the column density
distribution of \op absorption lines depends on the amplitude 
of the density fluctutions in the IGM.  
Croft et al  used the flux power
spectrum of QSO absorption spectra to constrain the DM power spectrum
on scales from $2h^{-1}$ to $12h^{-1}$ Mpc (Croft et al. 1998, Croft
et al. 1999, Weinberg et al. 1999).  Nusser and Haehnelt (1999)
(hereafter NH99) proposed an iterative procedure to obtain the real
space density and its  probability distribution function.  Here we
apply a modified version of this technique to two HIRES spectra of the
QSO Q1422+231. We infer the fluctuation amplitude of the DM density on
the effective Jeans scale and constrain the normalization constant of
the optical depth distribution which depends on the ionization rate 
$\Gamma_{\rm phot}$, the baryon density $\Omega_{\rm bar }$ and 
the Hubble constant ($ {\cal A }\propto \Omega_{\rm bar} ^2 h^3/\Gamma_{\rm
phot}$).  Section 2 describes our modified technique to extract 
moments of the probability distribution (PDF) of the density and 
presents tests on mock spectra.
In section 3 we present the application to the data. Section 4
discusses implications for the cosmological density parameter.  In
section 5 we relate our work to previous work and give an outlook.
Section 6 contains our conclusions.

\section{Theoretical Background}  

\subsection{Basic equations}

The optical depth in redshift space due to  resonant  \op scattering 
is related to the \hi\ density along the line of sight 
(LOS) in real space by 
\begin{equation}
\tau(w)=  \sigma_{0} \;\frac{c}{H(z)}\;\int_{-\infty}^{\infty} \dhi (x) 
{\cal H}[w-x-\vp(x) , b(x)]  
\dd x,  \\
\nonumber
\label{tau}
\end{equation}
where $\sigma_{0}$ is the effective cross section for resonant line
scattering, $H(z)$ is the Hubble constant at redshift $z$, $x$ is the
real space coordinate (in $\kms$), \calH is the Voigt profile
normalized such that $\int{\cal H }\, \dd x =1 $, $v_{\rm p}(x)$ is
the LOS peculiar velocity, and $b(x)$ is the Doppler parameter due to
thermal/turbulent broadening.  The absorption features in the 
\op\  forest are mainly produced by regions of moderate densities 
where photoheating is the dominant heating source and shock heating 
is not important.  Hydrogen in the IGM is highly ionized 
(Gunn \& Peterson 1965, Scheuer 1965 ) and  the photoionization 
equilibrium in the expanding IGM establishes a tight correlation 
between neutral and total hydrogen density. This relation can be
approximated by a power law  $\nhi \propto n_{_{\rm H}}^\alpha$, where
the parameter $\alpha$ depends on the reionization history.
The possible range for $\alpha$ is is $1.56 \la \alpha \la 2$ with a value  
close to $2 $ just after reionization, and 
decreasing at later times (Hui \& Gnedin 1997).  
Numerical simulations have shown that the gas density traces the 
fluctuations of the DM density  on scales larger than the Jeans length, so that
$ \dhi = \dhhi \left(\frac{\rho_{_{\rm DM}}(\vx)}{
{\bar \rho_{_{\rm DM}}}}\right)^\alpha$
, where $\dhhi$ is the 
\hi\ density at mean dark matter density.
In this relation $ {\rho_{_{\rm DM}}(\vx)}$ is the
dark matter density smoothed on the Jeans length scale
below which thermal pressure becomes important.
The Jeans length in the linear regime  
is given by (comoving),
\begin{eqnarray} 
x_{_{\rm J,lin}} &=& \frac {2\pi c_{\rm s}}{\sqrt{4\pi G \hat \rho}} \,
(1+z) \nonumber \\ 
& \approx & 1.3 
\left ( \frac{\Omega h^2}{0.125} \right )^{-1/2} 
\left ( \frac{\hat T}{1.5 \times 10^4\K} \right )^{1/2} 
\left ( \frac{1.5}{1 + (2-\alpha)/0.7} \right )^{1/2} 
\left ( \frac{1+z}{4} \right )^{-1/2} \; \Mpc,
\end{eqnarray} 
where $c_{\rm s}$ is the sound speed. 
However, in the non-linear regime gas can collapse to  scales
smaller than this and the Jeans scale becomes a somewhat ambiguous quantity. 
Here we define  an effective non-linear Jeans length, $x_{_{\rm J}}$, as 
the width of a kernel of the form $[1+ (k x_{_{\rm J}}/2\pi)^2]^{-2}$,
such that the rms fluctuation amplitude of  $ {\rho_{_{\rm DM}}(\vx)}$
is the same as that of the unsmoothed dark matter density  
filtered with this kernel (see section 4 for details).  
On scales larger than the effective Jeans length,   equation
(\ref{tau}) can be written as
\begin{equation}
\tau(w)=  {\cal A} (z)  \int_{-\infty}^{\infty}  
\left(\frac{\rho_{_{\rm DM}}(\vx)}{{\bar \rho_{_{\rm DM}}}}\right)^\alpha
{\cal H} [w-x-\vp (x), b(x)] \dd x ,
\label{tauw}
\end{equation}
with
\begin{eqnarray}
{\cal A} (z)&=& \sigma_{0}  \frac{c}{H(z)} \; \dhhi  \nonumber \\
&\approx&   0.61 \; 
\left (\frac{300 \kms \Mpc^{-1}}{H(z)}\right ) \;
\left (\frac{\Omega_{\rm bar}h^2}{0.02}\right )^{2} \;
\left ( \frac{\Gamma_{\rm phot}}{10^{-12}\s^{-1}} \right)^{-1}\; 
\left ( \frac{\hat T}{1.5\times 10^{4}\K} \right )^{-0.7} \;
\left ( \frac{1+z}{4} \right )^{6} , \nonumber \\
\label{tauf}
\end{eqnarray}
where $\Omega_{\rm bar}$ is  the baryonic density in terms of the 
critical density and $\Gamma_{\rm phot}$ is
the photoionization rate per hydrogen atom. The Doppler parameter in
the last equation depends on $\nhi$ as $b\propto \dhi^{1-\alpha/2}$.

NH99 have presented a direct Lucy-type iterative scheme to recover 
the optical depth and the corresponding mass and velocity fields 
in the LOS from the normalized flux,
$F=\exp(-\tau)$. They showed that the density field can  
be successfully recovered  below a  threshold value above which 
the corresponding  flux  saturates. 
NH99 further used mock spectra extracted from an
N-body simulation to show that the moments of the mass density can 
be reasonably well estimated by  fitting a log-normal PDF to the low
density tail of  the PDF of the  recovered density. In the next section we 
present a modified method for estimating the moments of the density 
distribution.

\subsection{Moments of the density PDF  from the 
recovered local optical depth} 

NH99 provided an estimate for the quantity 
\begin{equation}
\ttau(x)\equiv {\cal A} \left[\frac{\rho(x)}{{\bar \rho}}\right]^{\alpha} ,
\label{lod}
\end{equation}
which we  term the local optical depth. It is related to the 
observed optical depth $\tau$ by a convolution with a Voigt
profile as described in equation  (\ref{tau}). 
The recovered $\ttau$ is a good approximation to the true
field only in regions with $\ttau $ less than a certain value.
For large optical depths, the recovered $\ttau$ typically 
underestimates the true field. 
We  define  a  truncated local optical depth 
$\ttaut$ as  $\ttaut=\ttau $ for $ \ttau<\ttaum$, and $\ttaut=0$ otherwise.
The corresponding truncated moments of 
$\ttau_{\rm t}$ can be written in terms of the
density PDF as,
\begin{equation}
<\ttau^n_{\rm t}>= 
{\cal A}^n \int_{-\infty}^{\delta_{\rm c}} 
\left(1+\delta\right)^{n \alpha} P(\delta) \dd \delta,
\label{trunc}
\end{equation}
where $\delta=\rho/{\bar \rho} -1$ is the density contrast and
$\delta_{\rm c}=\left(\ttaum/{\cal A}\right)^{1/\alpha}-1$.
We further define  $\nu = [{\rm ln}(\rho/\bar \rho)-\mu_1]/\mu_2$,
where $\mu_1$ and $\mu_2$ are the average and $rms$ values of ${\rm
ln}\left(1+\delta\right)$. 
NH99 had also shown that the PDF of the DM density smoothed 
on the scale relevant for the \op forest (the effective 
Jeans scale) can  be reasonably well approximated by a log-normal distribution (Bi \& Davidsen 1997, cf. Sheth 1998 and  
Gaztanaga \& Croft 1998 for different forms
of the PDF).
For  a log-normal density distribution, 
$P(\nu)=\exp(-\nu^2/2)/\sqrt{2 \pi}$,  the truncated moments in 
(\ref{trunc})   can be written as,
\begin{equation}
<\ttau^n_{\rm t}>=
\frac{{\cal A}^n}{2} \exp\left(\frac{1}{2}n^2 \alpha^2 \mu_2^2 + n \alpha \mu_1\right)
\left[1+ {\rm erf}\left(\frac{\num - n \alpha \mu_2}{\sqrt{2}} \right) \right]
\label{taumom}
\end{equation}
where $\num$ is the value of $\nu$ corresponding to $\delta_{\rm c}$. 
By  expressing $\nu$ in terms of $\ttau$ in (\ref{taumom}) we find
\begin{equation}
<\ttau^n_{\rm t}>=
\frac{1}{2} \exp\left(\frac{1}{2}n^2 \alpha^2 \mu_2^2 - n \alpha \mu_2^2/2
+ n\ln {\cal A}    \right)
\left[1+ {\rm erf}\left
(\frac{\ln \ttaum - n \alpha^2 \mu_2^2 -\ln {\cal A} + \alpha \mu_2^2/2}
{\alpha \mu_2 \sqrt{2}} \right) \right] .
\label{taumomfin}
\end{equation}
We have here used the relation $\mu_1 =-\mu_2^2/2$ which follows 
from the condition  $<\delta>=0$ for the log-normal.
The moments of the truncated optical depth distribution depend on four
parameters ${\cal A}$ , $\mu_2$, $\alpha$ and $\ttaum$.  The parameter
$\ttaum$ is chosen   such that for $\ttau <\ttaum$ 
the local optical depth does not suffer from the biases introduced 
in saturated regions. As apparent from equation (\ref{taumomfin}) there are two
basic degeneracies leaving us with two independent parameters: 
\begin{equation}
{\cal B}\equiv \ln {\cal A} - \alpha \mu_2^2/2 , \qquad \qquad 
{\cal C}\equiv\alpha \mu_2.
\end{equation}
In terms of these two parameters, the moments of $\ttaut$ can finally 
be written as
\begin{equation}
<\ttau^n_{\rm t}>=
\frac{1}{2}\; \exp\left(\frac{1}{2}n^2 {\cal C}^2 + n {\cal B} \right)
\;
\left[1+ {\rm erf}\left
(\frac{\ln \ttaum - n {\cal C}^2 - {\cal B}}
{ \sqrt{2} {\cal C}} \right) \right].
\label{taumomfin}
\end{equation}
The first two moments, $<\ttau_{\rm t}>$ and  $<\ttau^2_{\rm t}>$,
are therefore sufficient to determine the parameters $\cal B$ and
$\cal C$. From these we can then infer the $rms$ fluctuation 
amplitude of the DM density and the normalization constant 
of the optical depth \cA\ as discussed in section 3.3. 
The use of truncated moments of the local optical depth
has a number of advantages. It is numerically more stable than
fitting a parametric form of the PDF to the distribution of the
recovered density. It allows  to introduce a cut-off in 
the form of an optical depth instead of the density 
which depends on the a priori unknown value  of \cA .
Finally, it  shows the basic degeneracies between $\alpha$, 
$\mu_2$ and \cA.

\begin{figure}
\centering
\mbox{\psfig{figure=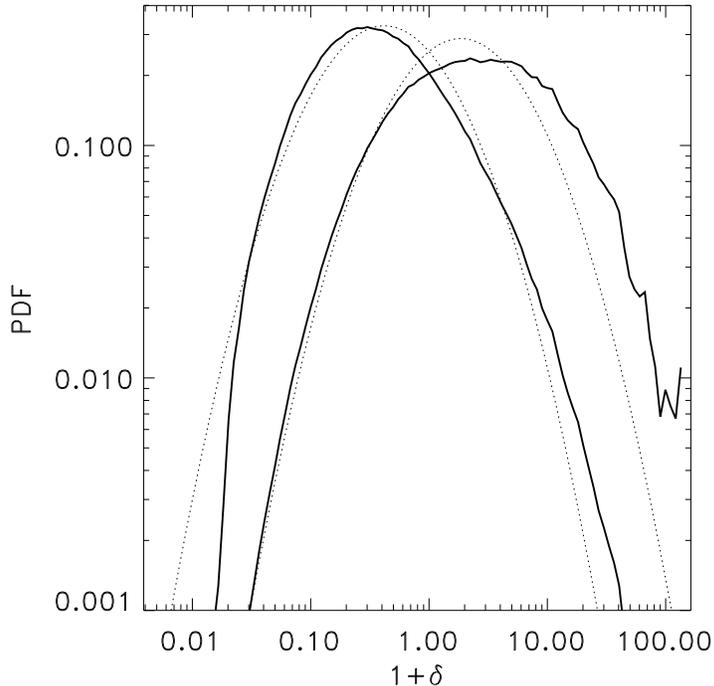,height=4.0in}}
\caption{The curves on the right are the PDFs of  ${\rm ln}(1+\delta)$
for an Edgeworth expansion around a log-normal distribution 
with $\mu_3=0.4$. 
The curves on the left show  $(1+\delta)P[{\rm ln}(1+\delta)]$.
The solid curves are  the average 
of  70 PDFs from which the mock spectra of Q1422 were generated.
The dotted lines correspond to a log-normal PDF 
with the same value of $\mu_2$.}
\label{fig:pdf_rho}
\end{figure}

In reality the PDF of the mass density smoothed 
on an effective Jeans scale is expected to be somewhat 
more positively skewed than a log-normal.  
The relations derived above can  easily be generalized to 
other forms  of the density PDF. 
We were, however, not able to constrain the additional skewness 
parameter of an Edgeworth expansion. The same was found by 
NH99.  We therefore  used a log-normal PDF for the analysis of 
the data. We assess  the possible biases  that arise in 
the determination of \cC\  and \cB\ using mock spectra in 
the next section.

\section{Application to the spectrum of Q1422+231} 

\subsection{The data} 

We had two spectra of Q1422+231 at $z=3.62$ available for our
analysis.  The spectra were observed with the HIRES spectrograph by
\retal\ (1997) and \cetal\ (1996) on the Keck I telescope.  The
resolution of the spectra is $\approx 8 \kms $ and the pixel size is
$2.5\kms$. We used the wavelength region from 4742 \AA (Ly$\beta$ at
the redshift of the QSO) to 5618 \AA (Ly$\alpha$ at the redshift of
the QSO) to avoid confusion due to an overlap of intervening
Ly$\alpha$ and Ly$\beta$ absorption.  This corresponds to a redshift
range from $z=2.9$ to $z=3.62$ and is equivalent to $240\mpc$
(comoving) in an Einstein-de Sitter Universe.  The two spectra were
continuum-fitted independently by the observational groups.  The
differences in our results should thus give an idea of the error due
to continuum-fitting.

\subsection{Error estimates and tests of the procedure 
using mock spectra}

Two sources of errors are important for our analysis: first the
intrinsic error caused by the inversion procedure; second the cosmic
variance error resulting from the use of only one quasar spectrum with
limited redshift range.  We quantify these errors and possible
systematic biases by analyzing mock spectra. The mock spectra were
produced using a modified version of the method suggested by Bi et
al. (1992) [see also Bi \& Davidsen (1997)].  The mock spectra have
the same redshift range, pixel size, resolution and S/N as the
observed spectra of Q1422.  We generated random realizations of a
Gaussian density field with a linear CDM-like power spectrum.  We then
assumed a parametric form for the PDF of the non-linear density and
obtained the non-linear density field along the LOS from the linear
field by a rank-ordered mapping (Weinberg 1992).  The corresponding
non-linear peculiar velocity, $v_{\rm pec}$, was computed as described
in NH99.  The optical depth was calculated using equation (\ref{tauw})
with $\alpha=1.7$ and $b = 16 \nhi^{0.15} \kms$.  The factor \cA\ was
chosen such that the mean flux decrement in the mock spectrum matches
the observed value. Finally, instrumental broadening as well as photon
and pixel (read out) noise was added.

Based on numerical simulations, we expect the true PDF to be more
positively skewed than a log-normal.   
Our method for estimating \cB\ and \cC\ assumes, however, 
that the  PDF is log-normal. It is
therefore important to test the sensitivity of the estimated moments
to the shape of the true PDF.  We generated mock
spectra for  a log-normal and for PDFs described by an Edgeworth
expansion around the log-normal,
\begin{equation}
P(\nu)=\left(2\pi\right)^{-1/2} \exp\left(-\nu^2/2 \right)
\left[1+\frac{\mu_2 \mu_3}{3!} \left(\nu^3- 3\nu\right) \right] ,
\label{edge}
\end{equation}
where the skewness parameter is defined as $\mu_3=<\nu^3/\mu_2>$.
N-body simulations with Gaussian initial conditions have shown this 
parameter to be  positive, independent of the detailed shape of the 
initial power spectrum (Colombi 1993, NH99). We will present 
detailed results for two different PDFs 
with  $(\mu_2,\mu_3=1.3,0.0)$  and $(1.0,0.4)$, but we have also
tested the technique with $(0.9,0.6)$.  
In figure \ref{fig:pdf_rho} we show representations of the PDF 
with $\mu_3=0.4$.  The deviations of the
PDF of $\left(1+\delta\right)$ from a log-normal distribution with the same
$\mu_2$ are significant only for $\delta \gsim 2$. 
Since only the low density part of 
the PDF is relevant for the truncated moments, 
the method is expected  to produce reasonable  
estimates  of $\mu_2$ even if the  true PDF is somewhat skewed relative to 
log-normal.  
We have generated  mock spectra for both PDFs with and without 
peculiar velocities to test the effect of redshift distortions. 
For each case 70 mock spectra (i.e. 70 random realizations 
for the same density power spectrum) were analyzed.

\begin{figure}
\centering
\mbox{\psfig{figure=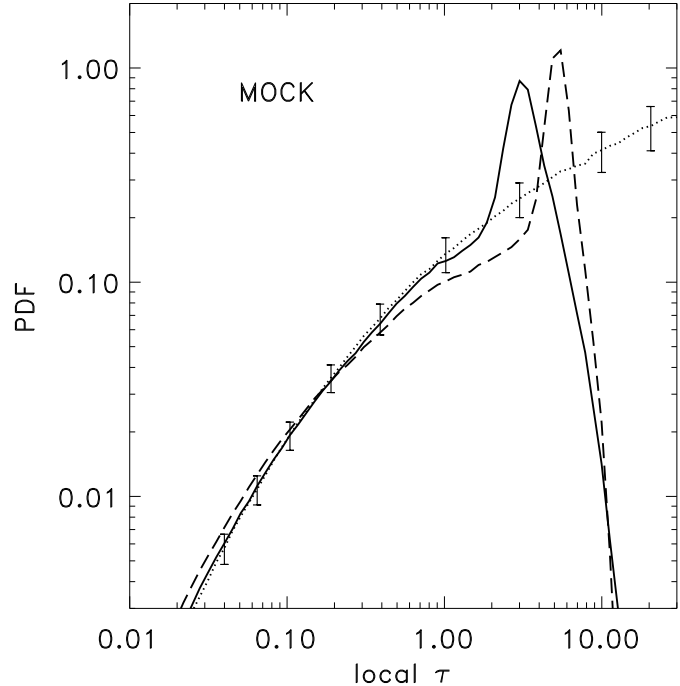,height=4.0in}}
\caption{
Curves of $\ttau P[{\rm ln}(\ttau)]$  for the recovered and true
local optical depth for mock spectra generated with $\mu_3=0.4$. 
The solid and dashed curves are for the recovered $ \ttau$ in real and 
in redshift space. The dotted line correspond to the 
mean of  the true PDFs 
of the 70 mock spectra. The error bars 
represent the cosmic variance error  ($rms$ values of the deviations 
of the mock spectra from the mean PDF).}
\label{fig:pdf_tau_mock}
\end{figure}

We have also modified the inversion procedure by introducing a power law
relation between the Doppler parameter and the local 
optical depth, $b=b_0 \ttau^{1-\alpha/2}$.
This has significantly speeded up the convergence
of the inversion. To  determine $b_0$ 
we define the function $\chi^2=
\sum_{\rm pixels}\left(F_i^{\rm obs} - F_i^{\rm
rec}\right)^2/\sigma_i^2$, where $\sigma_i$ is 
the error  and  $F_i^{\rm obs}$ the observed flux 
in pixel $i$. $F^{\rm rec}_i$ is the flux  which corresponds
the recovered local optical depth. We choose $b_0$ to be the largest value
which gives  $\chi^2/{\rm pixel}=1$ after a certain number of
iterations. 
\begin{figure}
\centering
\mbox{\psfig{figure=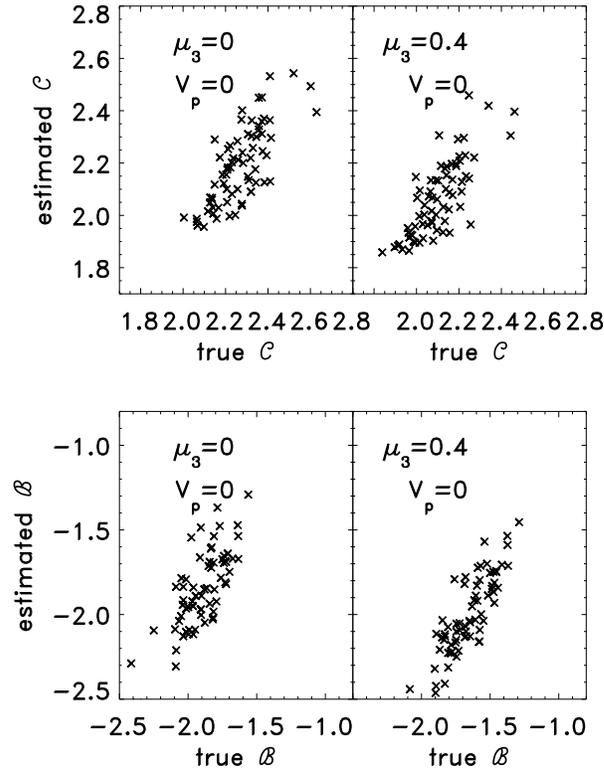,height=4.0in}}
\caption{The estimated vs true \cC\ and \cB\ for 
70 different random realizations of  mock 
spectra generated with $v_{\rm pec}=0$. }
\label{fig:parms_nv}
\end{figure}
\begin{figure}
\centering
\mbox{\psfig{figure=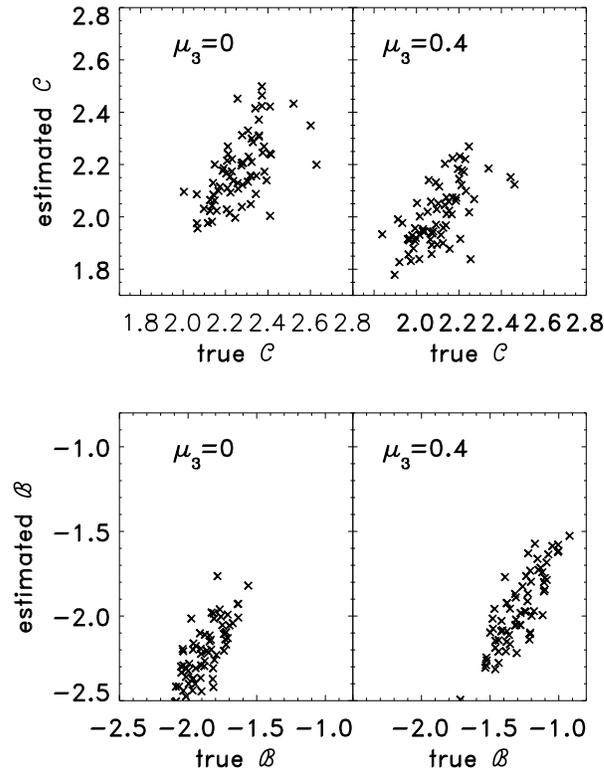,height=4.0in}}
\caption{The same as the previous figure but for spectra 
generated with non-vanishing peculiar velocities.}
\label{fig:parms}
\end{figure}

In Fig. \ref{fig:pdf_tau_mock} the average $\ttau P[{\rm ln}(\ttau)]$
of the 
recovered local optical depth of 70 mock spectra 
is shown (for the case with $\mu_3=0.4$) in real (solid curve ) 
and redshift space (dashed curve). 
The dotted line represents the average true PDF. The error bars are 
$1\sigma$ cosmic variance errors calculated from the 70 random
realizations. The spikes in the PDFs of the recovered local optical
depth are due to the systematic underestimate in saturated regions.  
Figure \ref{fig:pdf_tau_mock} demonstrates that we have successfully 
corrected for redshift distortions  in regions of low optical depths.
This correction shifts the location of the prominent spike to lower
$\ttau$ because the  redshift distortions tend to enhance the real 
space density contrast.  We use the location of the spike 
to choose the cut-off $\ttaum$ for the truncated moments.  

We have calculated  the first two moments $<\ttaut> $ and 
$<\ttaut^2>$ for three cutoff values, $\ttaum=1.0$, 
$1.5$ and $2.0$.  The parameters \cB\ and \cC\ were determined
by solving equations (\ref{taumomfin}) with Broyden's method 
(e.g. Press et al 1992). Figures (\ref{fig:parms_nv}) and
(\ref{fig:parms}) show the true {\it vs} the recovered values of the
parameters \cB\ and \cC\  for four sets of 70 
mock spectra (two input PDFs, with and without peculiar velocities).

The tight correlation between the true and recovered values 
demonstrates robustness of the  method. The estimated values  of \cC\
are unbiased in all four cases  while \cB\ is 
systematically underestimated (except in the log-normal case with
peculiar  velocities set to zero).
We  found  similar results for the mock  spectra  
produced with($\mu_2=0.9$ , $\mu_3=0.6$) 

\begin{table}
\caption{True and estimated parameters for the mock q1422+231 spectra
generated with a log-normal PDF of the mass density. Listed are estimated
parameters from mock spectra with and without peculiar velocities
$(v_{\rm pec} =0)$. The first entry shows the parameters estimated after
correcting for redshift distortions, the second before the correction
and the third the parameters from spectra generated with $v_{\rm pec}=0$.  The
first and second rows are $({\cal C} , {\cal B})$ and the $1\sigma$
error due to the the inversion.
The fourth entry lists the true values of the parameters and 
the $1\sigma$ error due to cosmic variance.  
The transformation from \cC\ and \cB\ to \cA\
is given by $\ln {\cal A}={\cal B} +{\cal
C}^2/2/\alpha$ and  requires a value for $\alpha$.}
\begin{tabular}{l|c|c|c|} \hline
&$\ttaum=1.0$ & $\ttaum=1.5 $& $\ttaum=2.0$\\ 
&(\cC,\cB)&(\cC,\cB)& (\cC,\cB)\\ \hline
Est. ({\rm real space})&$(2.27,-2.04) $&$ (2.18,-2.22)$&(2.16,-2.23) \\ 
         $\qquad {\rm inversion\ error}$    &$ (0.1,0.0.22) $&$  (0.10,0.14)$&(0.11,0.11) \\ \hline
Est. ({\rm redshift space})&$(2.25,-2.25) $&$ (2.23,-2.68)$&(2.23,-2.68) \\ 
        $\qquad{\rm inversion\ error}$    &$ (0.12,0.15) $&$  (0.11,0.12)$&(0.12,0.11) \\ \hline

Est. ($v_{\rm pec}=0$)&$(2.23,-1.70) $&$ (2.20,-1.81)$&(2.19,-1.85) \\ 
         $\qquad{\rm inversion\ error}$  &$ (0.08,0.25) $&$  (0.08,0.18)$&(0.10,0.15) \\ \hline

True  &$(2.27,-1.89)$& &  \\
           $\qquad{{\rm cosmic\ variance}}$    &(0.12,0.16)& &  \\ \hline

\end{tabular}
\label{tab:mock_lg}
\end{table}

\begin{table}
\caption{
The same as the previous table but for mock spectra generated 
using a skewed PDF with $\mu_3=0.4$. Notice that there are two entries 
for true values in this table. 
This is because 
different values of \cA\ are needed  
to match the observed flux decrement for mock spectra generated
with and without peculiar velocities
if $\mu_3 \ne 0$ .}
\begin{tabular}{l|c|c|c|} \hline
&$\ttaum=1.0$ &$\ttaum=1.5 $& $\ttaum=2.0$\\ 
&(\cC,\cB)&(\cC,\cB)& (\cC,\cB)\\ \hline
Est. (real space)&$(2.07,-1.83) $&$ (2.02,-1.95)$&(2.00,-1.97) \\ 
       $\qquad {\rm inversion\ error}$    &$ (0.09,0.18) $&$  (0.10,0.15)$&(0.10,0.13) \\ \hline
Est. (redshift space)&$(2.05,-2.26) $&$ (2.03,-2.29)$&(2.03,-2.29) \\ 
     $\qquad {\rm inversion\ error}$    &$ (0.10,0.16) $&$  (0.11,0.13)$&(0.10,0.10) \\ \hline

Est. ($v_{\rm pec}=0$)&$(2.09,-1.95) $&$ (2.07,-2.00)$&(2.06,-2.02) \\ 
       $\qquad  {\rm inversion error}$  &$ (0.08,0.19) $&$  (0.08,0.15)$&(0.09,0.13) \\ \hline

True ($v_{\rm pec}\ne 0$) &$(2.08,-1.30)$& &  \\
           $\qquad{\rm  cosmic\ variance}$    &(0.13,0.19)& &  \\ \hline
True ($v_{\rm pec} = 0$) &$(2.08,-1.67)$  & &  \\
            $\qquad  {\rm cosmic\ variance}$   & (0.13,0.19) &  & \\ \hline
\end{tabular}
\label{tab:mock_lge}
\end{table}

In tables (\ref{tab:mock_lg}) and (\ref{tab:mock_lge}) we list 
the mean values of $({\cal C}, {\cal B})$ and the estimated 
errors determined from the  four sets of 70 mock spectra. 
The intrinsic random error introduced in the inversion method is 
the $rms$ value of the differences between  estimated and 
true values from each set of
the 70 realisations.  The cosmic variance errors are
estimated as the $rms$ values of the true parameters.  For both the
skewed and the log-normal PDFs the estimated and true values of \cC\
agree well within the $1\sigma$ errors.  Redshift distortions only
moderately affect the estimated \cC\ despite  substantial
differences between the  PDFs (see figure
\ref{fig:pdf_tau_mock}). The estimates for  the three
different cutoff values are consistent within the errors, indicating a weak
dependence on $\ttau_{\rm c}$.

\subsection{Estimated parameters from Q1422}

\begin{figure}
\centering
\mbox{\psfig{figure=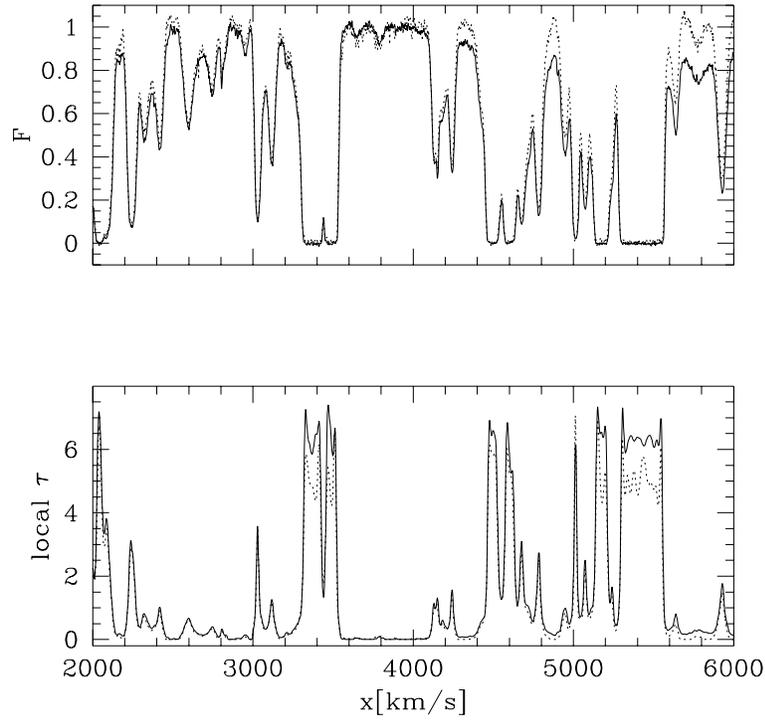,height=4.0in}}
\caption{ The normalized fluxes ({\it top}) and recovered 
local optical depths in redshift space ({\it bottom}) for a fraction 
of the two spectra of Q1422.  The solid and dotted lines correspond to
the spectra observed and continuum fitted by \retal\ and \cetal , 
respectively. }
\label{fig:tau_real}
\end{figure}

In the top panel of figure \ref{fig:tau_real} we plot the  
normalized flux of the two spectra of Q1422 obtained independently by  
Rauch et al. (solid line) and Songaila \& Cowie 
(dotted). There is a good general agreement between the two.
 As we will see later,  the discrepancies caused by 
 the different continuum fits have a moderate  
 effect on the  estimated parameters.
In the bottom panel of  figure \ref{fig:tau_real} we plot 
the recovered local optical depth in redshift space.  
The results for both spectra agree very well for $\ttau \la 3$. 
Both curves have a  cutoff in saturated regions where 
the flux goes to zero. The differences at $\ttau >3$ will 
not affect our estimates of \cB\ and \cC\ because of the truncation 
at smaller values of $\ttau$. 

\begin{figure}
\centering
\mbox{\psfig{figure=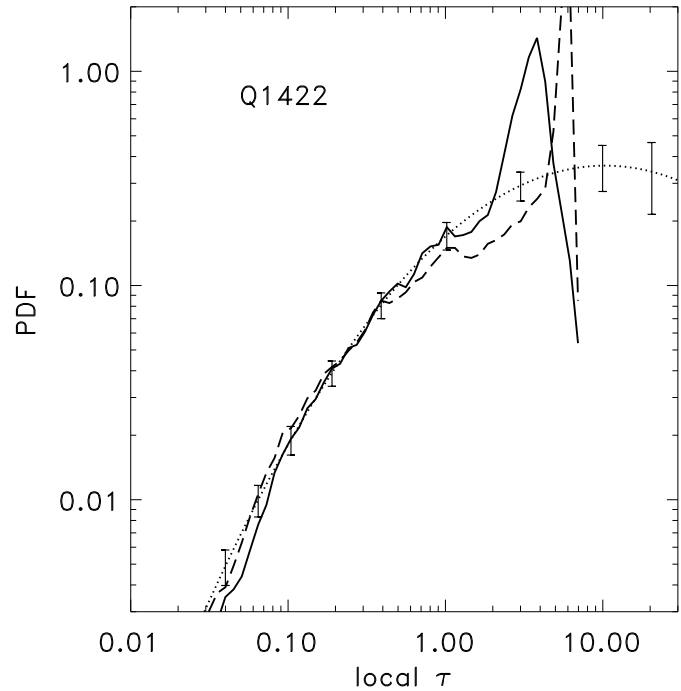,height=4.0in}}
\caption{ Curves of $\ttau P[{\rm ln}(\ttau)]$ for the local optical
depth in real (solid) and redshift space (dashed) recovered from the
\retal\ spectrum.  The dotted curve corresponds to the log-normal PDF
obtained from the estimated parameters $({\cal C}, {\cal
B})=(1.8,-1.61)$. The error bars are taken from the figure 2.}
\label{fig:pdf_tau_real}
\end{figure}

Figure (\ref{fig:pdf_tau_real}) shows $ \ttau P[{\rm ln}(\ttau)]$
for  the recovered $\ttau$
from the \retal\ spectrum, in real (solid line) and in redshift space
(dashed).  The dotted line is for the log-normal  with parameters
estimated from $\ttau$ in real space.  There is  reasonable agreement
between the recovered PDF and the log-normal for low 
optical depth implying that the true PDF cannot be very different 
from a log-normal  for  low densities.

In table (\ref{tab:real}) we list the estimated 
values of \cC\ and \cB\ and the value of  \cA\ 
inferred for $\alpha =1.7$.
Because of the strong redshift dependence of \cA\
(eq. \ref{tauf}), we present results from the first half, 
the second half and the total length of both spectra.
We have used a variable Doppler parameter  $b=b_0\ttau^{1-\alpha/2} 
$with $\alpha=1.7$.  A value of  $b_0=15{\rm  km/s}$ 
matches the $\chi^2$ criterion described in the previous
section.  
As discussed in section 3.2. our inversion procedure 
works well for cut-off values  $\tau_{\rm c}$ up to 2.   
For  $\tau_{\rm c}=2$ we obtain $({\cal C},{\cal B})=(1.8,-1.61)  $ 
and  $({\cal C},{\cal B})=(2.09,-1.61) $   for 
the Rauch et al and the  Songaila \& Cowie  spectrum, 
respectively. 
As discussed in section 3.2. the
value of \cB\  is expected  to be biased low. 

To assign errors  to the estimates of \cC\ and \cB\ we draw on 
the analysis of the mock spectra in the previous section.
We take  the maximum error of all sets of mock
spectra calculated for half the length of the observed spectrum
as a conservative error estimate.  This gives $\sigma_{\cal C}=0.27$. 
and $\sigma_{\cal B}=0.4$, respectively.  
The values of  \cC\ and \cB\  derived from the upper 
and lower half of the spectrum  are consistent within our 
estimated  errors if we account for the expected
redshift evolution of \cB .  The redshift evolution 
of the clustering amplitude and $\alpha$ and therefore that 
of \cC\ is expected to be small. 
The difference between the estimates obtained for the 
Rauch et al and the Songaila \& Cowie spectrum are also consistent 
within the estimated errors.

\begin{table}
\caption{Estimated parameters from the Rauch et al (R) and 
Songaila \& Cowie (SC) 
observations of the spectrum of q1422+231.
Listed are (\cC\, \cB , \cA) where \cA\ are computed from 
\cC\ and \cB\ using $ \ln( A) ={\cal B}+ {\cal C}^2/2/\alpha$ and assuming 
$\alpha=1.7$. The uncertainty in these parameters is estimated as the maximum
total error from the 4 sets of the 70 mock spectra (see text). 
The absolute $1\sigma $ on \cC and  \cB\ are 0.27 and   0.4.
}
\begin{tabular}{l|c|c|c|} \hline
&$\ttaum=1.0$&$\ttaum=1.5$&$\ttaum=2.0$ \\ 
&(\cC,\cB,\cA)&(\cC,\cB,\cA)& (\cC,\cB,\cA)\\ \hline
Low redshift half&&& \\ 
$\qquad{\rm real space}$, R &(1.83,-1.59,0.54)&(1.76,-1.74,0.43)&(1.70,-1.82,0.38)\\ 
$\qquad{\rm real\ space}$, SC &(2.21,-1.09,1.44)&(2.09,-1.78,0.60)&(2.05,-1.87,0.53)\\
$\qquad{\rm redshift\ space}$, R &(1.82,-1.91,0.39)&(1.79,-1.90,0.38)&(1.71,-1.97,0.32)\\
$\qquad{\rm redshift\ space}$, SC &(2.35,-1.71,0.92)&(2.22,-2.03,0.56)&(2.15,-2.132,.46)\\\hline

High redshift half&&&\\ 
$\qquad {\rm real\ space}$, R &(1.89,-0.89,1.17)&(1.89,-1.32,0.76)&(1.87,-1.35,0.72)\\
$\qquad {\rm real\ space}$, SC &(2.06,-0.63,1.86)&(2.12,-1.26,1.06)&(2.10,-1.34,0.95)\\
$\qquad {\rm redshift\ space}$, R &(1.99,-1.35,0.83)&(1.97,-1.44,0.74)&(1.89,-1.62,0.56)\\
$\qquad {\rm redshift\ space}$, SC &(2.29,-0.95,1.81)&(2.26,-1.55,0.95)&(2.126,-1.84,0.60)\\\hline
All& && \\ 
$\qquad {\rm real\ space}$, R &(1.89,-1.26,0.81)&(1.84,-1.55,0.57)&(1.80,-1.61,0.51)\\
$\qquad {\rm real\ space}$, SC &(2.14,-0.84,1.66)&(2.13,-1.52,0.83)&(2.09,-1.61,0.72)\\
$\qquad {\rm redshift\ space}$, R &(1.92,-1.66,0.56)&(1.89,-1.70,0.52)&(1.81,-1.82,0.42)\\
$\qquad {\rm redshift\ space}$, SC &(2.35,-1.33,1.34)&(2.26,-1.79,0.75)&(2.14,-1.98,0.53)\\\hline
\end{tabular}
\label{tab:real}
\end{table}

For a given value of $\alpha$ we can now infer 
values of $\sigma$ and \cA\  from our estimates of 
\cC\ and \cB .  The  values of \cA\  given in table (\ref{tab:real}) 
are derived using 
\begin{equation}
{\cal A}= \exp {[{\cal B} +{\cal C}^2/2\alpha]}. 
\end{equation}
Since \cB\ is systematically biased low,
the values of \cA\  in the table should be considered  as  
lower limits.  Assuming that the true  \cA\ is  biased low by a 
factor of 2 as suggested by our analysis of the mock 
spectra  with $\mu_3 = 0.4$ 
we estimate ${\cal A}(z\approx 3.25)   \approx 1.2\pm 0.7$
($\sigma_{\ln {\cal A}}=\sqrt{\left(\sigma_{\cal B}^2 + \sigma_{\cal
C} {\cal C}/\alpha \right)} \approx 0.65$ for $\alpha=1.7$). 

We now discuss the inferred value of $\sigma=<\delta^2>^{1/2}$. 
For a PDF described by an Edgeworth 
expansion (equation \ref{edge}),
 $\sigma$ and $\mu_2 =<(\log[1+ \delta])^2>^{1/2}$    are related  by 
\begin{equation}\sigma^2=\left( 1+{\mu_2^4 \mu_3}/{3!} 
\right)^{-1}\exp\left(\mu_2^2\right)-1  .
\label{sigmu}
\end{equation}
We have demonstrated in the previous section 
section that the  estimates of \cC\ are unbiased even
for true PDFs with rather large  skewness parameter $\mu_3$. 
The value of $\mu_2={\cal C} /\alpha$ is therefore also unbiased. 
Hence by  assuming $\mu_3=0$ we
somewhat overestimate  $\sigma$.
For the values of $\mu_2$ and $\mu_3$ measured in numerical  
simulations (NH99) we expect this bias to be less than 15\%,   
significantly  smaller than  the uncertainty in the value of $\alpha$. 
We therefore use the relation  
\begin{equation}
\sigma^2={\exp{[({\cal C}/\alpha)^2]}-1}. 
\end{equation} 
Note that the small  bias introduced here 
will also somewhat lower our estimates for $\Omega$ in the next 
section.  Possible values of $\alpha$ range between  1.56 and 2
but $\alpha$ is probably about  1.7 at $z\sim 3$ for 
plausible reionization histories (Hui \& Gnedin 1997,
Abel \& Haehnelt 1999).  The estimated value of ${\cal C} = 1.8\pm 0.27$
for the Rauch et al. spectrum translates into
$\sigma(\alpha=2.0)=1.11\pm 0.2$,$ \sigma(\alpha=1.7)= 1.44\pm 0.3
$, and  $ \sigma(\alpha=1.56) =1.67\pm 0.4$. The values for the 
Songaila \& Cowie  spectrum are 20 percent higher.

\section{Comparison with the mass fluctuations amplitude predicted by
CDM-like cosmogonies: implications for $\Omega$}

There is a whole family of CDM-like  power spectra which is
consistent with a variety of observational constraints on the DM
clustering amplitude.  If we normalize these power spectra to match the
local space density of galaxy clusters
(e.g., Eke, Cole \& Frenk 1996) then, for a given
$\Omega$, we can predict the fluctuation amplitude on the scale and
redshift relevant to the estimate of $\sigma$ in the previous
section.  
In the following we obtain constraints on $\Omega$
by comparing our measurement of $\sigma$ with such model predictions.

\begin{figure}
\centering
\begin{sideways}
{\psfig{figure=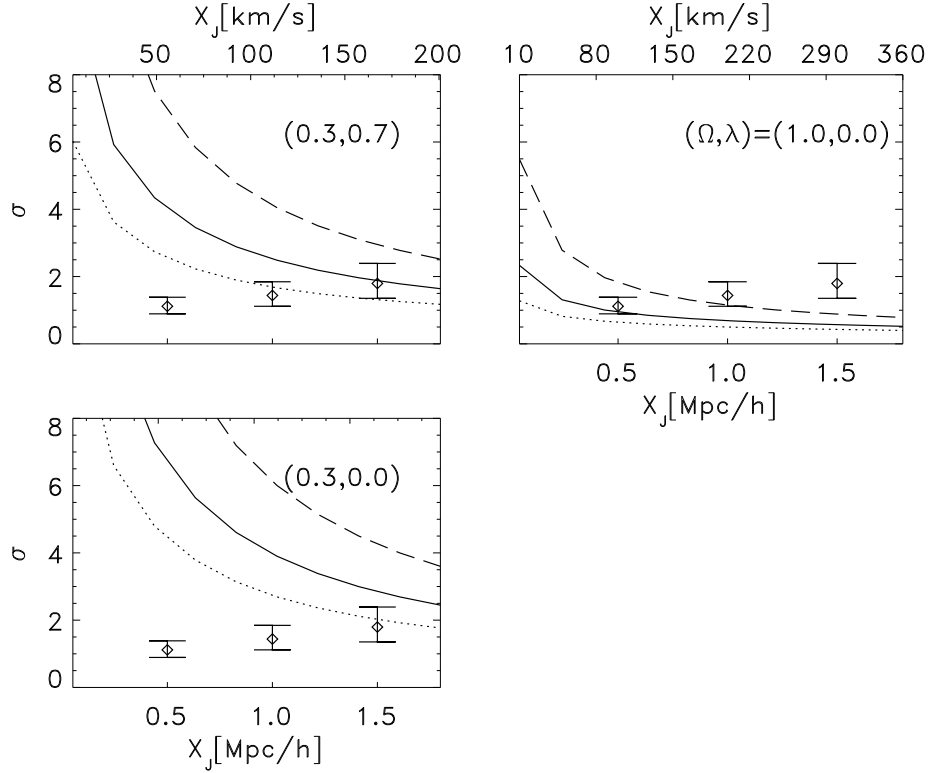,height=6.0in}}
\end{sideways}
\caption{The non-linear $rms$ value of density fluctuations as a function
of the effective Jeans length scale. The solid, dotted and 
dashed curves correspond to $\Gamma=0.21$, $0.1$ and $0.5$,
respectively.  The three points with error bars are the 
values estimated from the Rauch et al spectrum of Q1422 
for $\alpha=1.5$, $1.7$ and $2.0$ (from high to low). 
The abscissa of the data points are arbitrary and the errors
are $1\sigma$ and include the error due to cosmic variance and 
the inversion.}
\label{fig:modcomp}
\end{figure}

We start with the {\em linear} mass power spectrum  
as given by Bond, Eftsathiou, Frenk \& White (1993) 
\begin{equation}
P(k)=\frac{ {\mathrm A} k}{\left[       
1+\left( a k + (b k)^{3/2} + (c k)^{2} \right)^\nu
        \right]^{2/\nu}},
\label{pk}
\end{equation}
where the wavenumber $k$ is in units of $1/(\mpc)$, $\mathrm A$ is a
normalizing factor, $a=6.4\Gamma^{-1} \mpc$, $b=3.0 \Gamma^{-1} \mpc$,
$c=1.7\Gamma^{-1} \mpc$, and $\nu=1.13$.  For standard cold dark
matter models the constant $\Gamma$ is identified with $\Omega
{\mathrm h}$.  The space density of present-day galaxy
clusters is reproduced  if $\mathrm A$ is chosen such that 
the linear $rms$ value, $\sigma_8$, of the density fluctuations in 
a sphere of radius $8h^{-1} \mpc$ is given by 
\begin{equation}
\sigma_8=\left(0.52\pm 0.04\right) {\Omega}^{-0.52+0.13\Omega},\; 
\sigma_8= \left(0.52\pm 0.04\right) {\Omega}^{-0.46+0.1\Omega} ,
\label{sig81}
\end{equation}  
for flat and open models, respectively
(Eke, Cole \& Frenk 1996).
We then use an analytic  transformation 
from the linear to the {\em non-linear} power spectrum
(Peacock \& Dodds 1996; see also  Jain, Mo \& White 1995).

In section 2 we introduced an effective Jeans scale $x_{_{\rm J}}$ 
such that the fluctuation amplitude of the DM  density field 
smoothed on  this scale can be computed 
from the non-linear power spectrum as  
\begin{equation} 
\sigma^2(x_{_{\rm J}}) =\frac{1}{\left(2\pi\right)^3} 
\int_0^{\infty} \dd k^3 P_{\mathrm nl}(k)
\left[ 1+\left (\frac{k x_{_{\rm J}}}{2\pi}\right )^2 \right]^{-2}. 
\end{equation} 

Even though   $x_{_{\rm J}}$  should be related  to the 
linear Jeans scale, the exact choice of $x_{_{\rm J}}$
is not obvious  and we have left   $x_{_{\rm J}}$  as a free 
parameter. The nonlinear $\sigma$ depends
then on the shape parameter $\Gamma$, the background density, 
$\Omega$, the cosmological constant, $\lambda$,
and the effective Jeans  scale, $x_{_{\rm J}}$.  In figure
(\ref{fig:modcomp}) we show the model prediction for $\sigma$ at
$z=3.25$ as a function of the effective Jeans scale $x_{_{\rm J}}$ for
different cosmological parameters.  The value of $\sigma$
inferred from the Rauch et al. spectrum  
is shown as the open symbols with error bars for 
$\alpha =1.5$, $1.7$ and $2.0$.
Three factors contribute to 
the rather strong $\Omega$ dependence of
the predicted $\sigma$: 
the power spectrum normalization, the time evolution of the 
fluctuation amplitude and the relation between length and
velocity scale.  There is a similar but somewhat weaker 
dependence on $\lambda$. 

\begin{figure}
\centering
\begin{sideways}
{\psfig{figure=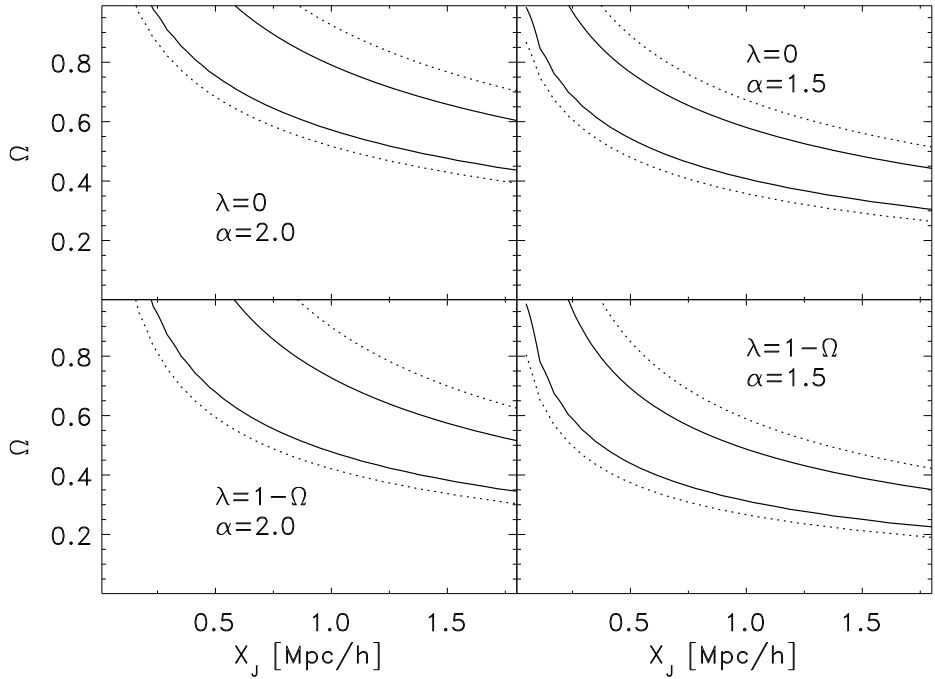,height=6.0in}}
\end{sideways}
\caption{Contours of $\chi^2=1$ (solid) and $\chi^2=3$ (dotted) as
functions of $\Omega$ and $x_{_{\rm J}}$ for two values of $\alpha$
for open and flat models, as indicated in each panel.  The
$\chi^2$ function quantifies the deviation 
of the measured clustering  amplitude  from that 
predicted by the $\Gamma=0.21$ model at
$z=3.25$.  The error includes cosmic variance, the error 
due to the inversion and the uncertainty in the normalization of the
linear power spectrum.}
\label{fig:modchi_all}
\end{figure}

For a given $\alpha $, we define the function $\chi^2= \left[{\cal
C}/\alpha-\ln(1+\sigma^2)/2\right]^2/\sigma^2_{\rm err}$ where the
$\sigma_{\rm err}$ is the total $1\sigma$ error including the
uncertainty in the power spectrum normalization (\ref{sig81}).  Figure
(\ref{fig:modchi_all}) is a contour plot of $\chi^2$ as function of
$\Omega$ and  $x_{_{\rm J}}$ for flat ($\Omega+\lambda=1$)
and open models ($\lambda=0$). 
If we  assume  $\Gamma=0.21 $ as inferred from the
shape of the power spectrum of the observed galaxy distribution at
$z=0$ (Peacock \& Dodds 1994) our estimate for 
$\Omega$ can be expressed 
in a convenient parametric  form as follows: 
 $\Omega= 0.61 
\left(\alpha/1.7\right)^{1.3}\left(x_{_{\rm J}}/0.62\right)^{-0.6}$ for a flat
universe, and $\Omega= 0.91
\left(\alpha/1.7\right)^{1.3}\left(x_{_{\rm J}}/0.62\right)^{-0.7}$ for an open
universe without a cosmological constant. The errors on $\Omega$ are 25\%. 
The inferred values from the \cetal\ spectrum have the same parametric
form but are lower by 15\%.

\section{Discussion}

\subsection{Comparison to previous work}

\vspace{0.25cm}
\noindent
{\it The value of \cA} 

Rauch et al. have used several HIRES spectra including the one 
of Q1422+231 used here to infer the value of 
$\mu = (\Omega_{\rm bar} h^{2}/ 0.0125)^2
(H(z)/100\kms\Mpc^{-1})^{-1} (\Gamma_{\mathrm phot}/10^{-12}\s^{-1})^{-1}$ 
by comparing the flux decrement distribution of the  observed  
spectra to those of mock spectra calculated from hydrodynamical 
simulations.  This analysis was basically 
equivalent to a determination of the  optical  depth
constant  \cA.  The two quantities are related 
as   
\begin{equation} 
{\cal A}(z) \approx 0.71\;  \mu (z)\;  
(\hat T/1.5\times 10^{4}\K)^{-0.7}\;  ((1+z)/4)^6.  
\end{equation} 
Rauch et al obtained  $\mu(z\approx 3.25)  \approx  0.6$ and $  1.5$  
for mock spectra produced from a SCDM model 
(Hernquist et al. 1996) and $\Lambda$CDM model (Cen et al. 1994) . 
The temperatures at $z=2$ were $\hat T = 5.6 \times  10^{3} \K$ 
and $\hat T = 1.1 \times 10^{4} \K$.  
This  corresponds to  ${\cal{A}}(z=3.25) \sim  1.2 $ and 
and $2.0$, respectively. 
No correction for the temperature evolution in the simulations has 
been made.
This is  in reasonable agreement with our estimate  of 
${\cal A} \sim 1.2  \pm 0.7$ but we  would like to stress 
that our value includes a rather  large correction 
of a factor two for the bias in \cB\ which we have inferred from our 
analysis of mock spectra produced with a PDF with $\mu_3 = 0.4$. 
We therefore do not think that we 
can currently improve the constraints of Rauch et al. on the 
baryon density and the flux of ionizing photons. The fact that 
we obtain a consistent result with a completely different 
procedure  is nevertheless very encouraging. 
Our value  is larger 
than  the  ${\cal A}(z\approx 3.25)\sim 0.4$ in the fiducial model of 
Bi \& Davidsen  ($\alpha = 1.77$).

\vspace{0.25cm}
\noindent
{\it The value of $\sigma$}

Gnedin and Hui (1998) used mock spectra calculated from DM simulations
to investigate the influence of the fluctuation amplitude on the slope
of the column density distribution of QSO absorption lines.  They got
reasonable agreement with the observed slope if the linear fluctuation
amplitude at half the linear Jeans scale for $T=10^4\K$ 
was about 2. This is larger than our value for the {\it non-linear} $\sigma$. 
This  indicates that the effective Jeans scale is  larger than 
 half the linear Jeans scale for $T=10^4\K$.
Bi\& Davidsen have assumed $\sigma = 1.95$ 
in their fiducial model significantly larger than our value 
$\sigma(\alpha=1.77) = 1.34$ for the same value of $\alpha$. 

Croft et al. (1998, 1999) have compared the flux power
spectrum of observed absorption spectra to infer the DM power spectrum
on scales between $2 \mpc$ and $12 \mpc$ (comoving, $\Omega =1$ ) by
comparison with mock spectra of DM simulations which were calibrated
using hydrodynamical simulations.  Note that Croft et al. define their
length scales to be a factor $2\pi$ larger than we do in our
definition of the Jeans length.  Croft et al obtain $\Delta = k^3 P(k)
/2\pi^2 \approx 0.57 \pm 0.3$ at $z \approx 2.5$, where $\Delta$ is
the rms fluctuation amplitude of the DM density contributed by a
logarithmic interval of $k$. This measurement was at a scale which
corresponds to $x\approx 12 (\Omega h^2/0.125)^{-0.5} \Mpc$.
Weinberg et al. (1999) then used the constraints 
on the DM power spectrum 
obtained by Croft et al. (1999) to  constrain  the value of $\Omega$ 
with similar  assumptions for the normalization and the shape of the
power spectrum  as we have made here. 
For $\Gamma =0.2$ Weinberg et al. 1999 obtain  
$\Omega \approx 0.34 \pm 0.1$ ($\lambda =1-\Omega$) and $\Omega \approx 
0.46 \pm 0.1$ ($\lambda =0$). We obtain somewhat higher values 
unless the effective Jeans scale is larger than the linear 
Jeans scale (for a temperature of $15000\K$) by about a factor two.

\subsection{Future work}

There are a number of possible improvements on the analysis 
presented here. As discussed in section 
3.4 the true PDF of the  density is expected to be somewhat 
positively skewed relative to log-normal, 
but we were not able to constrain the skewness  parameter,
$\mu_3$. To make progress on this 
issue an improved model of the PDF which e.g. relates  
$\mu_3$ to other parameters would be needed. 
There are attempts to determine the value of $\alpha$  from the 
correlation of column density and the lower cut-off of the Doppler 
parameter  distribution (Schaye et al. 1999). Such a determination 
will obviously improve our constraints on 
\cA\ and  $\sigma$. Currently we probe the density PDF 
up  to a rather low  overdensity of $\delta \sim 2$. 
This could be extended to higher densities by incorporating 
higher order Lyman series line into the inversion scheme. 
This will tighten the  constraints on the density PDF 
and also extend the redshift range of each spectrum 
available for the inversion.  The main source of uncertainty  in our  
constraints on $\Omega$  is the effective Jeans lenght scale $x_{_{\rm J}}$.  
A calibration of $x_{_{\rm J}}$ by numerical simulations should help 
to tighten the constraints on $\Omega$. It should also 
be possible to extract information on the shape of the power spectrum 
on  small scales from the correlation function of the truncated 
real-space optical depth.  We estimated the error of $\sigma$ due 
to cosmic variance to be about 20\% for one spectrum like Q1422.
This error could be reduced to 5\% by applying the method 
to 20 quasar spectra. Note that the error estimate was based on mock
spectra generated with a SCDM power spectrum. It might be
different for other power spectra. It would obviously also be interesting 
to probe the redshift evolution of the clustering amplitude.

\section{Conclusions}

We have recovered the  real space optical depth distribution along 
the LOS to the QSO Q1422+231 from two different HIRES spectra of 
Q1422+231 using the inversion technique proposed by Nusser 
\& Haehnelt (1999). Despite the two spectra having   different 
continuum fits,  the difference in the recovered optical depth
distribution is small. The recovered optical depth distribution 
is similar to that expected for a log-normal PDF 
of the density. 

We have used a new technique to estimate the 
fluctuation amplitude of the gas density 
and the optical depth constant from the truncated optical 
depth distribution. Assuming that the PDF of the gas density 
is log-normal we showed that the moments of the distribution depend 
on two parameters ${\cal B}= \ln{[{\cal{A}}/(1+\sigma)^{\alpha/2}}]$   and  ${\cal C}= \alpha 
\sqrt{\ln{[\sigma^{2}}+1]}$.  
This demonstrates that there are  fundamental degeneracies between 
fluctuation amplitude of the gas, the power law index $\alpha$,
and the optical depth constant \cA\ for a log-normal PDF. 
These degeneracies will  be only very weakly broken  if the  
true  PDF is somewhat positively skewed 
relative to  log-normal  as expected from numerical simulations. 

Using mock spectra we have shown that robust estimates for the
parameters \cB\ and \cC\ can be obtained.  The mean of the 
recovered \cC\ is unbiased 
with respect to the true value. The value of \cB\, however, is systematically
biased low if the true PDF is positively skewed relative to 
log-normal. Our estimates of \cB\ and \cC\ translate 
into  ${\cal A} \sim 1.2 \pm 0.7$  and $\sigma  \sim 1.44\pm
0.27$ if we assume $\alpha=1.7$, the most likely value for 
plausible reionization histories. The error estimates come from our 
analysis of  the mock spectra. If the IGM indeed traces the DM in the
way suggested by simulations this is the first measurement of 
the fluctuation amplitude on  an effective Jeans scale 
which corresponds to about  two hundred
kpc(comoving). The fluctuation amplitude of the DM density 
on such small scales is of particular interest for the 
question when the epoch of  reionization occurs. Our rather 
low estimate for the value of $\sigma$ strongly argues for a late 
epoch of reionization. 

Finally, we have compared our measurement of the fluctuation amplitude 
to predictions  by CDM-like cosmogonies with a power spectrum shape 
parameter suggested by the clustering behavior of galaxies, and  normalized 
to match the observed space density of present-day clusters
of galaxies.  The fluctuation amplitude of such models 
depends rather strongly on Omega.  We obtain  
$\Omega= 0.61\left( \alpha/1.7\right)^{1.3}
\left(x_{_{\rm J}}/0.62\right)^{-0.6}$ ($\lambda = 1- \Omega$)  
and $\Omega= 0.91 \left( \alpha/1.7\right)^{1.3}
\left(x_{_{\rm J}}/0.62\right)^{-0.7}$ ($\lambda=0$). 
If we assume that the effective Jeans scale is close to the linear
Jeans scale (at  15000K) then the measured fluctuation amplitude is
larger than that  predicted for an ($\Omega = 0.3$, $\lambda = 0.7$)  
Universe by a factor of about two. 
This argues either for a larger value of $\Omega$ and/or 
a large value  of the effective Jeans length $x_{_{\rm J}}$. 

\section{Acknowledgements}
We are grateful to L. Cowie, M. Rauch, W. Sargent and
A. Songaila for letting  us 
use their HIRES spectra of Q1422 for the analysis 
presented here. We would like to thank T. Theuns and S. White
for comments on the manuscript.

\protect
\end{document}